\documentclass[11pt]{article}
\usepackage{geometry}                
\usepackage{graphicx}
\usepackage{amssymb}
\usepackage{epstopdf}
\title{Objective Methods for Assessing Models for Wildfire Spread}
\author{Jeff Picka}
\date{}
\begin{document}
\maketitle

\begin{abstract}
Models for wildfires must be stochastic if their ability to represent wildfires is to be objectively assessed. The need for models to be stochastic emerges naturally from the physics of the fire, and methods for assessing fit are constructed to exploit information found in the time  evolution of the burn region. 
\end{abstract}

\section{Introduction}

Management of wildfires requires the development of real-time forecasting methods for the prediction of their evolution. Development of these models requires acknowledging the difficulties associated with data acquisition, data processing, and modeling of wildfire events. At present, many predictive models are deterministic, and produce one forecast for every set of observed fire conditions. When a single deterministic prediction is made, two major questions arise: what does that single prediction represent, and how reliable is it? There is no way to answer either of these questions unless the deterministic model is replaced by a model which satisfies three conditions. 

\begin{enumerate}
\item The model predicts only a single aspect of the fire rather than predicting many diverse aspects of the fire.
\item The model must be stochastic, and make predictions in the form of estimated probability distributions or probability maps.
\item The fit of any model must be assessed by statistical methods after it has been calibrated.
\end{enumerate}
These conditions are imposed by our lack of ability to model the fire as a physical process, by our inability to observe the details of the internal structure of the fire, and by the difficulties associated with simulating sequences of regions of random shape and size. 

\subsection{Terminology}

A \textit{trajectory} is a sequence of regions of irregular shape and extent which describe the cumulative region burnt by a fire over time. Even if the prediction problem does not directly involve the burned region, it will almost certainly involve predicting a response statistic which is defined by the shape and size of that region (e.g. cumulative value of timber burned, expenditure required to extinguish the fire).  

A \textit{stochastic model} is a collection of stochastic processes whose outcomes may represent fire trajectories. The \textit{parameters} of the stochastic model are collections of numbers which index the stochastic processes in the model. A \textit{fitted (stochastic) model} is the one 
stochastic process in the model which is chosen to best represent  the fire for which a forecast needs to be made. The fitted stochastic model can be made to produce a sequence of realizations, each of which may represent a fire trajectory.

Since fire trajectories are unpredictable, the burn regions at any fixed time are unpredictable in shape, size, and orientation. At a fixed time, a sample of these shapes is a sample from a \textit{random set} process \cite{mol:2005}. The stochastic model for the fire spread is a time series of random set processes. 

\section{Stochastic Modeling of Fires}

Useful models for fires must be stochastic because of the nature of wildfires and our ability to model complex multiscale phenomena. The stochastic aspects of wildfires are unavoidable, and cannot be ignored in model construction, model fitting, or model assessment.

\subsection{Models for Multiscale Physical Phenomena}

The simplest version of a multiscale physical phenomenon consists of a collection of fundamental elements at one scale (the microscale)  which act collectively to produce a distinctive pattern of behaviour at a much larger scale of observation (the macroscale). At the microscale, the fundamental elements may or may not be observable. Even if they are observable, it may be difficult to model their behaviour. It is assumed that the macroscale phenomena are always observable.

In the ideal case, the behaviour of the fundamental elements can be used to derive the behaviour of the macroscale phenomena. The classical example of this is the Gibbs canonical ensemble model of an ideal gas, from which the basic results of classical thermodynamics for an ideal gas at equilibrium can be derived \cite{minlos}. If the state of an inert gas changes very slowly, then its physical behaviour can be modeled well by results derived from the Gibbs model. This model works well because the fundamental units are atoms of an inert gas, which have simple structure and which interact rarely. The simplicity of the microscale behaviour combined with the immense number of atoms involved results in macroscale behaviour which can be comprehensively described by a very small number of macroscopic state variables. Microscale simplicity also allows the derivation of models for the relationships between state variables, which do not need to include any consideration of macroscale fluctuations induced by microscale behaviour. 

Models for the steady-state laminar flow of Newtonian fluids and the steady-state conduction of heat through solids are also successful multiscale models \cite{bsl}. In these models, the mechanisms of energy transfer through small volumes is simple, and macroscale flow can be modeled using partial differential equations. In the case of heat transfer by convection or turbulent fluid flow, rigorous definition of macroscale properties from microscale behaviour is impossible, since there are no mathematical models that can faithfully represent the behaviour of the fundamental elements and their interactions. In other cases such as the modeling of the flow of powders \cite{kakalios}, the number of fundamental elements involved is not sufficient to eliminate fluctuations in properties at the macroscale.

In the modeling of mass, heat, and momentum  transfer, it is rare to be able to formulate a model which faithfully captures microscale phenomena and then derives from the microscale model a useful description of the limiting macroscale phenomena. Any model for macroscale behaviour depends on being able to summarize that behaviour with appropriately defined state variables, and then being able to construct a model which describes the relationships between the state variables. To be of use, this model must not be seriously inconsistent with the physics at the microscale, even if the microscale behaviour is unobserved and unmodelable. This approach can be successful if the model developer is modeling an inherently simple multiscale phenomenon, manages to develop a useful set of state variables, manages to find a model to relate them, and then finds that the resulting model produces predictions which are close enough to reality to solve the problem of interest. 

\subsection{Modeling of Wildfires}

Wildfires are not amenable to modeling by simple multiscale models which have deterministic macroscale behaviour.  
The combustion takes place in a vapour which forms above pyrolyzing fuel. The heat produced by the combustion drives the formation of more vapour, as well as slowly changing the geometry of the combustible materials from which the vapour is produced. The combustible material occurs in many different forms on many different length scales, and does not have homogenous structure. The heat evolved from combustion causes turbulent flow of air and of combusting and non-combusting fuel vapour. These turbulent flows are of sufficient strength to affect local weather. The modeling of any one of these phenomena is very difficult on a small scale. The violent and complex interactions of these processes cannot be observed except on the largest scales. 

The complexity of wildfires makes it impossible to imagine modeling them at the microscale and then establishing the limiting macroscale behaviour. The relatively small size of the macroscale relative to the microscale together with the complexity of interactions between the fundamental elements ensures that, no matter how much is known about the past history of that fire, there will be many possible future fire trajectories which are consistent with that history. Given what can be observed about a real fire, there is an ensemble of fire trajectories which are consistent with these observations. Given our inability to comprehensively model the fire, nature appears to choose at random one of the fire trajectories from the ensemble. Any model for a fire trajectory has to sample from that ensemble in the same way that the physical process would, and must yield a probabilistic rather than a deterministic prediction of future fire behaviour. 

Stochastic modeling of wildfire behaviour is possible, since the large-scale behaviour of the fires is more predictable than small-scale combustion phenomena.  Any model proposed will need to be stochastic, but there will not be any rigorous theoretical argument suggesting exactly what form this model should take. The trajectory of each real fire will be impossible to model, and so any model must be based on a random sample of some form of approximate trajectory. Construction of such models will require much creative programming and physical intuition. When constructing models, it would be unreasonable to assume that the first successful models will be useful in all prediction problems, or that the selection among possible approximated fire trajectories should be undertaken via a uniform distribution. Instead, it would be best to focus on one prediction problem at a time and then try to unify approaches once successful models have been found.

\subsection{The Risks Associated With Not Using Stochastic Models}

In many models now in use, the macroscale behaviour of a fire is simulated by a set of partial differential equations whose numerical solution produces a single approximate fire trajectory. This approach to modeling has many inherent problems which can be avoided by using stochastic models.

The most difficult problem associated with a single predictive trajectory is that it is not clear what that trajectory represents. If it is a single trajectory from the ensemble, then is it a typical trajectory or one with very unusual features when compared to other trajectories in the ensemble? Alternatively, the deterministic model may produce some form of average of many trajectories. This implies that at a fixed time, the burn region in the trajectory is an average of many such regions from the ensemble. Any region produced by a geometrical averaging process will be much simpler in shape than any single burn region from a trajectory in the ensemble. 

Even if the nature of the single prediction could be determined, in both cases the single prediction overstates the accuracy of the prediction procedure. The deterministic prediction can be used to produce graphics which can mislead a naive observer as to how uncertain the prediction really is, and gives no information whatsoever as to how typical that prediction is of others from ensemble. 

Given a single trajectory from any model, it is almost certain that the prediction will differ from the observed trajectory of the fire. If this difference is trivial to the eye, the model may be judged to be adequate in the modeling of that fire. If major differences are observed, it may be tempting to conclude that the model does not fit. To make such a judgment objectively, especially in a case where the ensemble contains a great variety of possible trajectories, it is necessary to make the judgment of model fit based on comparison between the observed trajectory and many simulated trajectories from the model. This process requires the use of a stochastic model, and is the subject of the next section. 

\subsection{Constructing a Stochastic Model}

If a deterministic model samples a single trajectory from the ensemble of physically viable trajectories, then it can be used as the basis of a stochastic model \cite{braun}. Stochastic elements can be added which represent lack of ability to clearly observe initial and boundary conditions, but they can also be added to compensate for aspects of fire propagation that are not fully understood. 

\section{Assessment of Model Fit}

Any stochastic model consists of a family of stochastic processes, only one of which is chosen to represent the fire. There is no guarantee that any method of calibration will produce a good choice, and so the fit of the chosen model must be assessed by statistical means. 

\subsection{Fitting a stochastic model}

Fitting a model involves finding a method to choose parameter values so that the fitted model can provide useful predictions of the evolution of a wildfire. This choice must be based on the past history of the fire.

The parameters can be divided into two classes. Physical parameters appear in those aspects of the model derived from physics, and take the form of physical constants such as the acceleration of gravity. These parameters are generally set using values established by physical experiments, often in contexts very different from wildfires. While these parameters are not generally allowed to be set by a model-fitting procedure, allowing these values to be fit may be useful in model assessment.

Calibration parameters are all parameters in the model which are not set by physical arguments. They may be physical parameters which cannot be observed or estimated, or they may be parameters which have no clear physical meaning but which make the model flexible enough to fit the data. They can be set by the guesses of experts who are attempting to make the model output look right, or by statistical means.  

Development of objective fitting procedures requires that the model predict only one specific aspect of the fire. Once the specific aspect is chosen, it is necessary to find a small set of response statistics which fully describes that aspect. Seeking a model which works in all predictive problems would require finding a small set of statistics which could describe or predict all possible responses. It is doubtful that any such universal set of predictive statistics exists.

Given a set of response statistics, a fitted stochastic model can be found by comparing the observed response to simulated responses. A large number of combinations of calibration parameters could be defined, and then a large sample of realizations generated from each stochastic process. From each sample, a mean response would be found. The fitted model would be the stochastic process whose mean response was closest to that of the observed fire. A response consisting of a time series of measurements would be best for use in this type of fitting procedure.

It is also possible to fit a model by varying the calibration parameters until the simulated responses look right. This method needs to be compared to objective methods since, in the hands of an expert, it may work well in some cases. This method may miss better parameter values, may be subject to conscious or unconscious bias, or may produce conclusions which are impossible to explain to others. 

\subsection{Assessing the fit of a fire spread model}

Once a fire spread model is fit, the adequacy of the model must be assessed by criteria appropriate to the prediction problem. This is necessary not only because all models are descriptive and speculative, but also because there is a severe risk that the fitting process will only match superficial similarities between the fire and the model realizations. 

Fit assessment for a fire spread model requires comparing a sample of simulated trajectories with the trajectory of the fire that they are modeling. If no evidence can be found of differences which could affect the response, then the model can be accepted as being useful. If evidence of differences are found, then these may suggest how the model can be improved. 

To assess the fit, the first stage is to summarize aspects of the fire trajectory using descriptive statistics.  These statistics may be based on the burn region at one time, or on a sequence of burn regions in time. They may be based on the shape of the burn region or on observable and quantifiable properties of the fire within it. 

For a descriptive statistic to summarize essential elements of the prediction problem, it is necessary for the response to depend on the statistic. This dependency can only be investigated by conducting statistical tests on fitted model output and then by attempting to identify associations. It would be better to investigate associations using fire data, but this is impossible unless the fire can be repeated many times under the same conditions. This classification process can reduce the number of statistics which are used to compare the model and the data. If a statistic is found not to affect the response, then there is no need for its values to be consistent for the model and the data. Since the classification is based on statistical inference, there is always a risk that a descriptive statistic will be misclassified. 

It is necessary to find a large number of descriptive statistics in any model assessment problem. Many descriptive statistics will be spatial, and will depend on the shape and size of regions. Statistics of this kind suffer from a lack of power, and may not be able to distinguish between spatial processes with visibly different realizations \cite{badsil}. There is also no general theory to suggest which statistics may be useful in finding differences between the response in the fire and the response in the fitted model.  

Finding descriptive statistics is a major challenge. Given expert concerns or past model failings, it is possible to quantify these in a similar manner to the development of response statistics. Using these statistics alone is not good enough, since they may not identify differences that have not been anticipated or observed in previous modeling efforts. Realizations must be summarized by statistics that can identify aspects of model failure that are difficult or impossible to see, even by experts. This requires developing a large library of statistics, all of which describe different aspects of the burn region. Analogues of second moment statistics, such as the pair correlation function, the nearest neighbour function, the empty space function, the $K-$function, and the 2-point correlation function \cite{stoy:1995} are inadequate for this purpose, since the stochastic model is not Gaussian. These statistics are very useful, but must be supplemented by other statistics including (possibly) statistics based on triangulations, statistics based on models for physical properties applied out of context \cite{picka:2005}, and statistics which have yet to be invented.

Given a set of descriptive statistics which affect the response, it is necessary to establish if they show any evidence that the model is wrong. This requires comparing a sample of a collection of descriptive statistic values from the observed fire with a sample which can be generated from many realizations of the fitted model. This comparison could be made with one descriptive statistic at a time, but better results are expected if small collections of statistics are compared. These comparisons are difficult to make reliably with the data from only a single fire, but this is unavoidable since each fire takes place under different conditions. Comparisons could be better made between data from controlled burns undertaken under fixed experimental conditions, but there is no guarantee that these burns would sample from ensembles similar to those associated with real fires. 

The assessment of fit will be subject to unavoidable errors since it is based on statistical tests. Investigations of association between descriptive statistics and the response will misclassify statistics. There will be no guarantee that the observed trajectory is typical of trajectories from physical process, and so differences between the observed and model trajectories may be found for a useful model. 
It is also possible that no evidence of poor fit will be found in models which are seriously wrong. This will happen if none of the statistics used to compare the model to the data are capable of identifying the difference. These errors are also unavoidable in fits by expert judgment.

\section{Conclusions}

Models which faithfully represent what is known and unknown about fire spread must be stochastic. The goal of a stochastic model is not to reproduce any one burn model exactly, but instead to produce simulated fire trajectories which approximate fire trajectories from the ensemble of possible fires. They need not capture every aspect of these fires, but they need to be able to capture the distribution of the response over the ensemble of real fires. The assessment of fit looks for evidence that the approximate trajectories are inconsistent with the one trajectory that is observed. A model will be useful if this general approach to fitting and assessment allows it to be predict outcomes successfully from many different fires.

The objective approach to model fitting presents a number of challenges which must be addressed before it can be usefully implemented. It requires the development of new statistics for the description of fire trajectories. It requires the development of a stochasticized version of a model like Prometheus \cite{tymstra} which can usefully be fit to data. It will be necessary to use historical data and simulations to determine what must be observed in order to predict fire behaviour and to assess the fit of fire models. It may require the acquisition or development of new equipment in order to gather the information on which predictions are based. 

Given the effort required, it may seem easier to continue fitting models by eye. This approach implicitly assumes that experts will not be influenced by any beliefs they may have about the fitness of model and that an expert will be able to see all possible ways in which the model could be wrong. If these assumptions are incorrect, then mistaken assumptions about fire behaviour could live on in flawed models for far longer than they should. Without a trajectory-based assessment of the quality of a fitted model, it will always be possible that a fitted model will fail to capture the physics of fire spread. 

The use of stochastic methods is also essential to making useful predictions about fire behaviour. If a model is purely deterministic, there is no way to determine if its single prediction is typical of those from the ensemble of possible fires or not. Also, there will no measure of uncertainty arising from the model that can be assigned to the prediction. This could lead to a loss of faith in modeling, due to failures in predictions for new fires. Deterministic predictions could also mislead non-experts into overconfidence about what is known and can be predicted about fire behaviour. Neither outcome would be beneficial for the development of useful models for fire spread. 

\section{Acknowledgments}

This work has been partly supported by NSERC but has mostly been supported by GEOIDE. 


\begin{thebibliography}{1}

\bibitem{mol:2005}
I.~Molchanov.
\newblock {\em Theory of Random Sets}.
\newblock Springer, 2005.

\bibitem{minlos}
R.A. Minlos.
\newblock {\em Introduction to Mathematical Statistical Mechanics}.
\newblock American Mathematical Society, 2000.

\bibitem{bsl}
R.B. Bird, W.E. Stewart, and E.N. Lightfoot.
\newblock {\em Transport Phenomena}.
\newblock Wiley, 1960.

\bibitem{kakalios}
J.~Kakalios.
\newblock Resource letter {GP}-1: {G}ranular physics or non-linear dynamics in
  a sandbox.
\newblock {\em Am. J. Phys.}, 73:8--22, 2003.

\bibitem{braun}
T.~Garcia, J.~Braun, R.~Bryce, and C.~Tymstra.
\newblock Smoothing and bootstrapping the {PROMETHEUS} fire growth model.
\newblock {\em Environmetrics}, 19:836--848, 2008.

\bibitem{badsil}
A.J. Baddeley and B.W. Silverman.
\newblock A cautionary example on the use of 2nd-order methods for analyzing
  point patterns.
\newblock {\em Biometrics}, 40(4):1089--1093, 1984.

\bibitem{stoy:1995}
D.~Stoyan, W.~S. Kendall, and J.~Mecke.
\newblock {\em Stochastic Geometry and Its Applications}.
\newblock John Wiley \& Sons, 2nd edition, 1995.

\bibitem{picka:2005}
J.D. Picka and T.D. Stewart.
\newblock A conduction-based statistic for description of orientation in
  anisotropic sphere packings.
\newblock In {\em Powders and Grains 2005 Proceedings}, pages 329--332.
  Balkema, 2005.

\bibitem{tymstra}
C.~Tymstra, R.W. Bryce, B.M. Wotton, and O.B. Armitage.
\newblock Development and structure of prometheus: the canadian wildland fire
  growth simulation model.
\newblock Technical Report NOR-X-417, Natural Resources Canada, Canadian
  Forestry Centre Edmonton, 2009.
\newblock forthcoming.

\end{thebibliography}

\end{document}